\documentclass[11pt]{article}

\usepackage{amsmath,amssymb,amsfonts,latexsym,graphicx}
\numberwithin{equation}{section}
\newcommand{\be}{\begin{equation}} \newcommand{\ee}{\end{equation}}
\newcommand{\bea}{\begin{eqnarray}} \newcommand{\eea}{\end{eqnarray}}
\newcommand{\field}[1]{\mathbb{#1}}

\title{
\begin{center}
 { \bf  A note on gauge theories coupled to gravity}
\end{center}
}

\begin{document}
\renewcommand{\thefootnote}{\fnsymbol{footnote}}
\rightline{RU-06-08}\rightline{SCIPP-06-07}
\begin{center}
 { \huge \bf  A note on Gauge Theories Coupled to\vskip0.2cm\ Gravity}

\end{center}\vskip0.4cm\
\centerline{\bf  Tom Banks$^{\star,\dagger,\S}$, Matt Johnson$^{\dagger}$ $\&$ Assaf Shomer$^{\star}$}
\vskip 0.5 cm
\centerline{\it
$\star.$ Santa Cruz Institute for Particle Physics}
\centerline {\it University of California, Santa Cruz, CA 95064}
\vskip 0.3 cm
\centerline{\it
$\dagger.$ Department of Physics}
\centerline {\it University of California, Santa Cruz, CA 95064 }
\vskip 0.3 cm
\centerline{\it
$\S.$ Department of Physics and NHETC}
\centerline {\it Rutgers University, Piscataway, NJ 08540 }
\vskip 0.4 cm
\centerline{\it
Email: banks,mjohnson,shomer@scipp.ucsc.edu}
\bigskip\bigskip\bigskip\bigskip

\begin{abstract}

We analyze the bound on gauge couplings $e\geq m/m_p$, suggested by
Arkani-Hamed et.al. We show this bound can be derived from simple
semi-classical considerations and holds in spacetime dimensions
greater than or equal to four. Non abelian gauge symmetries seem to
satisfy the bound in a trivial manner. We comment on the case of discrete
symmetries and close by performing some checks for
the bound in higher dimensions in the context of string theory.
\end{abstract}
\newpage

\setcounter{footnote}{0}
\renewcommand{\thefootnote}{\arabic{footnote}}

\section{Introduction}

In the context of string theory there are by now several examples of
consistent models of quantum gravity, but we are far from understanding
the full catalog of such models. It is therefore of interest
to formulate {\it general lessons} coming from the marriage of
gravity and quantum mechanics. An archetypal example is Hawking's
robust calculation of black hole (BH) evaporation~\cite{hawking}.
This calculation can be formulated in purely semiclassical terms,
and there are also cases where it can be derived as a rigorous
consequence of string theory. Another, somewhat less rigorous, idea
is the holographic principle~\cite{thooft, susskind, bousso}.

On a more solid footing is the general result that the only allowed
continuous global symmetries in quantum gravity are asymptotic gauge
transformations including asymptotic diffeomorphisms of space-times
with symmetric asymptotics. This statement can be derived from a
general semiclassical argument based on black holes, which will be
reviewed below and is supported by arguments from perturbative
string theory~\cite{tbetal} and from AdS/CFT \cite{witten}.

Recently, Arkani-Hamed et.al.~\cite{AMNV} (AMNV) suggested a general
bound governing any consistent theory of p-form gauge fields coupled
to gravity (the case $p=0$ was previously studied by Banks et.al in \cite{bdfg}). Their suggestion, which was motivated in part by
holography and in part by experience from string theory, can be
rephrased in several ways.
Take a model that includes GR in $3 + 1$ asymptotically flat
space-time dimensions, as well as a $U(1)$ gauge field with coupling
$e$.  In effective field theory, $e$ and Newton's constant $G$ are
independent. In particular, from the traditional point of view of
effective field theory, if we assume that $e$ is small enough so
that the Landau pole is above the Planck scale, we would generally
assume a cutoff of order $m_p$. Arkani-Hamed et.al. claim that
\begin{enumerate}
\item There has to be a light charged particle satisfying

\begin{equation}\label{bound}
m<e\cdot m_p.
\end{equation}

\item The effective gauge theory breaks down at a prematurely\footnote{Condition $2$ follows from condition $1$ e.g. by
considering the magnetic monopole as the light particle for the
magnetic interaction}$^{,}$\footnote{It is perhaps worthwhile
mentioning that while this suggestion seems surprising from the
perspective of effective field theory, it is expected (at least
qualitatively) within string theory. The reason is that both the
gravitational and the gauge interactions are governed by the string
coupling $g_s$. In perturbative string theory, the ``premature''
cutoff is the string scale itself.} low scale $\Lambda<e\cdot m_p$.
\end{enumerate}

We note that if condition $1$ is strictly satisfied as an
inequality, then extremal black-holes are kinematically able to
decay\footnote{If the extremal holes are BPS states then they are
marginally stable against decay. When the gauge charge in
question is in the SUSY algebra, the lightest BPS state will exactly
saturate the AMNV bound.}. Indeed, take an extremal BH in 4 dimensions satisfying
$GM=\sqrt{G}eQ$ where $M,Q$ are the mass and the {\it integral}
charge and $G\sim l_p^2=\kappa^2$ is Newton's constant. A
necessary condition for such a BH to decay is the existence of a
particle in the theory with a smaller mass/charge ratio\footnote{The
particle with smallest mass/charge ratio is exactly stable. Take a
particle of charge/mass ration $\gamma=M/Q$ which decays into a
bunch of particles of smaller charge $\sum q_i=Q$ and mass $\sum
m_i<M$ and assume none of them have smaller mass/charge ratio
$\gamma_i=m_i/q_i>\gamma$, then $\sum m_i=\sum \gamma_i q_i>\gamma
Q=M$  in contradiction.} which if normalized to $q=1$ gives
condition $1$.

As suggested in~\cite{AMNV}, condition $1$ written as
\begin{equation}\label{eb}
e>\frac{m}{m_p}
\end{equation} bounds the gauge coupling away from zero and therefore
is a {\it generalization} of the statement that there are no continuous global symmetries in a consistent
theory of quantum gravity. The logic behind this
statement is that if \ref{eb}\ is not satisfied then one can draw pathological conclusions such as
\begin{itemize}
 \item Charged BHs can posses entropies larger than their Bekenstein-Hawking entropy.
\item Charged BHs can decay to Planck size remnant with very large entropies. Such particles can violate entropy bounds and potentially
dominate the phase space of thermodynamic systems.
\end{itemize}

The purpose of this note is to fill in arguments left implicit by
AMNV and provide a semiclassical derivation of the AMNV bound from
the covariant entropy bound.  The paper is structured as follows. In
section \ref{GSym}\ we review semiclassical arguments against having
global symmetries\footnote{From now on, unless explicitly stated, we
only discuss continuous symmetries.} in a consistent theory of
quantum gravity. In section \ref{youone}\ we review the
Reissner-Nordstrom solution of Einstein-Maxwell theory and explain
our strategy for extending the argument of the previous section to
the case of gauge symmetries. In section \ref{disbh}\ we discuss the
 semi-classical loss of charge from a RN black-hole in a way that generalizes easily to higher dimensions.
 In section \ref{gengen}\ we present a semi-classical derivation of the AMNV bound in any space-time dimension
 bigger than or equal to 4. In section \ref{highdim}\ we make some checks of the validity of the AMNV bound
 in higher dimensions, within string theory. In section \ref{nags}\ we make some comments on the case of
 non-abelian gauge symmetries and discrete symmetries. Our conclusions are in section \ref{conc}.

\section{Global Symmetries in QG.}\label{GSym}

Let us first remind the reader of the black hole based argument that
there can be no global symmetries in a consistent theory of quantum
gravity.

\subsection{No-Hair}

A first indication of what is going on
comes from the {\it no hair} theorem. Local $U(1)$ symmetries obey a
form of Gauss's law which enables any observer {\it outside} the BH
horizon to determine its charge. On the other hand, if there were
also strictly Global symmetries then when a charged particle is
thrown into the BH, there is no way to determine this fact from the
outside. It thus looks like the charge was ``deleted'' in
contradiction to its conservation. However, at this level of
analysis, one can simply assign a charge to the BH and avoid this
difficulty.

\subsection{Hawking evaporation}

The real issue with global symmetries has to do with
Hawking evaporation and the problem of remnants. Since for a global
symmetry there is no associated gauge interaction, we can throw as
many charged particles as we want into a BH, and increase its charge
$Q$ to any value. The BH clearly can not radiate any appreciable
amount of charge in the Hawking radiation before $T_{Hawking}\geq
m$, where $m$ is the mass of the lightest charged particle.
Demanding that at the time when $M\leq m_p^2/m$ the BH still has
enough mass to be able to get rid of all its charge, we find
\begin{equation}
Q\leq ({m_p\over m})^2.
\end{equation}
This bound can be violated by taking $Q$ large enough.

In fact, the problem is worse. The Hawking radiation is thermal and
particles of opposite charge will be produced in equal numbers. The
BH will thus not get rid of any charge before becoming a Planck size
remnant. Such a situation leads to the following pathologies:

One can have a black hole of fixed mass $M$ and
arbitrarily\footnote{If the charge is not high enough, throw another
charged particle in and wait for the BH to evaporate the excess
mass.} high charge $Q$. Since an external observer cannot discern
the global charge, his micro-canonical ensemble of all states,
regardless of charge, with fixed energy inside a fixed area, would
lead us to to count microstates of all charge, and assign an entropy
at least of order $\log Q$ to the BH. This can be taken arbitrarily
high and contradict the Bekenstein-Hawking formula. If however, as
suggested by exact degeneracy counting in string theory, we accept
that the Bekenstein-Hawking entropy is a count of the number of
states of a black hole, such objects are ruled out, and global
continuous symmetries with them. Conversely, the assumption that the
BH entropy alone is a measure of the number of states, can be seen
as the requirement that all black holes can discharge (which in turn
implies the AMNV bound).

The problem is made worse by Hawking evaporation, which in a theory
with an unbounded global charge would lead to an infinite number of
Planck mass remnants. The charged Planck size remnants are particles
that can be confined to a small box and their entropy bounded by
something much smaller than the Bekenstein Hawking entropy of the
initial black hole.  Since these remnants can have any charge $Q$,
the entropy associated with the remnant in the box should be
infinite.

Alternatively, we can view the infinite degeneracy associated with
the global charge $Q$ as a variant on the species problem. Such
particles may dominate the computation of any low energy matrix
element of a product of two operators. This is not necessarily the
case if there is an additional form factor, analogous to the form
factor suppression of the production of monopoles in gauge theory.
Susskind~\cite{lennyremnant} appears to sidestep this objection by
considering the production of remnants in the thermal bath of
Rindler space, where the production of a species is fixed by its
mass alone. Said differently, one cannot make the existence of an
infinite number of stable low mass remnants consistent with the
assumption that Rindler or black hole horizons are thermal systems.

A crucial feature of all these arguments is the fact that the
hypothetical global charge is unobservable from the exterior, so
that measurements are made on ensembles of states with uncertain
charge $q$.   In some sense, this feature changes discontinuously
when we turn on a coupling between the conserved charge and a gauge
field.   However there is a clear sense in which there is not in
fact a discontinuity.   Measurements of charge are done by
performing scattering experiments and are ultimately limited by the
energy and momentum resolution of detectors.    As the gauge
coupling is taken to zero for fixed resolution, larger and larger
integer charges will become unobservable.   Roughly speaking, a
change of charge will have to be $o(1/e)$ in order to be detectable.
(Recall that for electromagnetism in the real world, ${1\over e}
\sim 3$).   In our discussion of entropy below, we will be referring
to ensembles with charge uncertainty of this order.

\section{Black-holes and $U(1)$ gauge symmetries.}\label{youone}

The Classical Einstein-Maxwell theory
\begin{equation}\label{em}
\mathcal{S}=\frac{1}{16\pi G}\int\sqrt{g}\mathcal{R}-\frac{1}{4e^2}\int\sqrt{g}\mathcal{F}^2
\end{equation} has the Reissner-Nordstrom (RN) solution of an electrically charged black hole in $4$ dimensions
\begin{equation}\label{rn}
\begin{split}
ds^2&=-f(r)dt^2+\frac{dr^2}{f(r)}+r^2 d\Omega_2^2,\qquad A=\frac{e^2Q}{r} \\
\rm{with}\quad  f(r)&=1-\frac{2GM}{r}+\frac{Ge^2Q^2}{r^2} \quad \rm{and}\quad \mathcal{F}=dA.
\end{split}
\end{equation} Here $M$ is the ADM mass and $Q$ is the integral charge quantized in units of $e$. The BH (outer) horizon lies at
\begin{equation}\label{arpl}
r_+=GM+\sqrt{(GM)^2-Ge^2Q^2}.
\end{equation} This solution is generally believed to make sense physically (i.e. obey Cosmic Censorship) only when
\begin{equation}\label{extb}
(GM)^2\geq Ge^2Q^2 \quad\Leftrightarrow\quad \boxed{M\geq eQ m_p}
\end{equation} where we have been cavalier about order one coefficients in
denoting $m_p\sim 1/\sqrt{G}$. Notice that the horizon is always of order $M$, namely
\begin{equation}\label{rp}
GM\leq r_+\leq 2GM.
\end{equation}

The extremal RN BH saturates the inequality in Eq.~\ref{extb}. This extremality bound can be understood physically by neglecting the effects of
gravity, and finding the energy stored in the electric field
resulting from bringing a charge $Q$ in from infinity
 to $r=r_+\sim GM$
\begin{equation}
E_Q=\frac{1}{e^2}\int_0^{r_+}\dfrac{(\frac{e^2Q}{\frac{4}{3}\pi
r_+^3})\frac{4}{3}\pi r^3\cdot (\frac{e^2Q}{\frac{4}{3}\pi
r_+^3})4\pi r^2dr}{r}\sim\frac{e^2Q^2}{GM}.
\end{equation} We now require that the actual BHs mass should at least account for this energy
\begin{equation}
M\geq \frac{e^2Q^2}{GM}\quad\Rightarrow\quad M\geq eQm_p,
\end{equation}
and thus recover Eq.~\ref{extb}.

The extremality bound Eq.~\ref{extb}\ applies for any non-zero gauge coupling $e$
and reflects the existence of repulsive gauge forces between
particles with the same sign of charge. The existence of this bound
clearly goes in the ``right direction'' of avoiding the problems
associated with global symmetries because one can not increase the
charge of the BH without also increasing its mass.

\subsection{The basic strategy}

To begin the argument for the AMNV bound, we assume the BH starts
with some mass and charge obeying the extremality bound
Eq.~\ref{extb}, and follow the spontaneous loss of charge by the BH.
We make no special assumptions about how this black hole was created
or the absolute value of its mass and charge. The discharge of the
black hole is exponentially suppressed below a certain threshold
that depends on the parameters of the BH. Below this threshold the
BH can always lose mass via Hawking radiation of massless particles
but will discharge very little. We then demand that at the threshold
the BH still has enough mass to be able to radiate away all of its
charge, namely, to account at least for the rest mass of $Q$ of the
lightest charged particles:
\begin{equation}\label{crit}
M_{discharge}\geq Q\cdot m_{light}.
\end{equation}
Even though we can track the evolution of the black hole reliably
only down to some mass scale of order $m_{p}$, the point here is
that energy and charge conservation allow us no conceivable way to
evaporate the black hole if Eq.~\ref{crit} is not satisfied.
This is a rephrasing of the criterion used in \cite{AMNV} that we
want to allow even extremal BHs to decay.

Otherwise, we run into essentially the same set of problems listed in section \ref{GSym}\ for global symmetries.
\begin{itemize}
\item Thinking in a grandcanonical ensemble, the energy band between $(M,M+\delta M)$ with $\delta M\ll M$ contains a huge number of black holes with the integral charge allowed by Eq.~\ref{extb} to be any number $Q=1,\dots,\frac{1}{e}$ leading to an entropy of order $\mathcal{S}\sim -\log e$. By taking the gauge coupling $e$ to be too small (e.g. take $e\sim 1/Q$) this entropy can be much bigger than the Bekenstein-Hawking entropy which is bounded (using Eq. \ref{extb}) by $\log (eQ)$.
\item We can have Planck size ($M \simeq m_{p}$) remnants with charge given by
any number $Q=1,\dots,\frac{1}{e}$, Again, the entropy one would associate with a system made up of putting such a particle in a box can be dialed to contradict any entropy bound by taking $e$ too small.
\item The species argument continues to work, where instead of an infinite number of species we only have a finite but very large $\sim 1/e$ number of species.
\end{itemize}

Notice that for very massive objects, where we can safely use the
classical Lagrangian \ref{em}, we can recover \ref{bound} by
insisting, as suggested in \cite{AMNV}, that extremal BHs can decay,
so the extremal mass should obey \ref{crit}, namely
\begin{equation}\label{potp} M_{ext}=Ge^2Q^2\geq
G^2(Qm)^2\quad\Rightarrow e>\frac{m}{m_p}.
\end{equation}
Following the strategy outlined in this section, we now show that
when we take into account the actual quantum effects of Schwinger
pair production and Hawking evaporation, responsible for any BH's
discharge (not necessarily an extremal one), we obtain the same
bound.

\section{Semi-classical discharge of Black Holes}\label{disbh}

Gibbons \cite{gibbons}\ gave an exact analysis, using Bogoliubov
transformations, for the spontaneous loss of charge by 4-dimensional
Reissner-Nordstrom black holes. In this section we summarize the
relevant results in a manner amenable to generalization
to higher spacetime dimensions.

Assume, as we do here, that there is a lightest charged particle of
mass $m$. The physical process responsible for discharging the hole
can be heuristically understood as follows. A pair is produced
outside the BH horizon and the member whose charge is opposite to
the BH is attracted and falls into the BH, while the other member of
the pair escapes to infinity.

It is useful to discuss two opposite regions of parameter
space, that of small/hot black holes where the discharge
is dominated by Hawking evaporation of charged particles and that of
large/cold black holes where the discharge is dominated by Schwinger
pair production~\footnote{The physical process of discharging the hole is really
the same in both cases in the sense that both correspond to the pair
production of particles of opposite charge.}.

\subsection{Small/Hot BHs - The Hawking process}\label{hot}
This regime is defined by
\begin{equation}
 GM\ll 1/m\quad\Leftrightarrow\quad T_{Hawking} \gg m
\end{equation} where the Hawking temperature is larger than the mass, $m$,
of the lightest charged particle so the latter will be thermally produced. In
contrast to the global case, where the thermal nature of the
radiation did not allow the hole to discharge, here the
electric field outside the horizon gives a ``chemical potential''
term in the Boltzman factor which is asymmetric between the charges
\begin{equation}
\mathcal{P}(m,\pm 1)\sim e^{-\frac{1}{T}(m\mp \frac{e^2Q}{r_+})}.
\end{equation} This favors the emission of a particle with the same
sign of charge as the BH. This is crucial for the discharge of the hole,
and kicks in when the contribution from the electric potential becomes
of order the rest mass $m$, namely
\begin{equation}
\frac{e^2Q}{r_+}\geq m.
\end{equation}

\subsection{Large/Cold BHs - The Schwinger process}\label{cold}

This regime is defined by
\begin{equation}
 GM\gg 1/m\quad\Leftrightarrow\quad T_{Hawking} \ll m
\end{equation} where the BH temperature is not large enough to produce
the lightest charged particle by the Hawking process discussed
in the previous section. Instead, the dominant mechanism for discharge is
Schwinger pair-production in a constant electric field\footnote{This
is the regime of parameters relevant for the decay of extremal BHs
which have zero temperature.}. While the field is not constant
everywhere, we can safely approximate the electric field outside the
horizon as almost constant for Large BHs. Gibbons' result in
this regime predicts the following rate for charge loss due to
pair-production
\begin{equation}
\frac{dQ}{dt}\sim e^{-\frac{\pi m^2 r_+^2}{eQ}}
\end{equation} which is exactly the dependence expected from the Schwinger process \cite{schwinger}. The hole starts to discharge appreciably when
\begin{equation}\label{schw}
\frac{m^2 r_+^2}{eQ}\leq 1.
\end{equation}
In fact, one can guess this dependence. Ask when the electric field
right outside the horizon has enough energy {\it in a volume of size
the Compton wavelength of the lightest charged particle} to account
for the creation of two of those particles from the vacuum. Since
the energy density in the electric field is given by
\begin{equation}\label{eef}
\epsilon=\frac{1}{e^2}\int d^4x \vec{E}^2
\end{equation}
and in this case $|\vec{E}|=\frac{e^2Q}{r^2}$ we can demand that
\begin{equation}
\frac{1}{e^2}\vec{E}^2\cdot\lambda_c^3\sim\frac{e^2Q^2}{r_+^4}\cdot\frac{1}{m^3}\geq
m
\end{equation} resulting in Eq.~\ref{schw}.

In the next section, we use these results to derive the AMNV bound in arbitrary spacetime dimensions.

\section{Deriving the AMNV bound in $N+1$ spacetime dimensions.}\label{gengen}

Following Meyers and Perry~\cite{myersperry}, the generalization of Reissner-Nordstrom BHs to higher dimensions is given by
\begin{equation}\label{grn}
\begin{split}
ds^2&=-f(r)dt^2+\frac{dr^2}{f(r)}+r^2 d\Omega_{N-1}^2, \\
f(r)&=1-\frac{16\pi GM}{(N-1)A_{N-1}r^{N-2}}+\frac{8\pi Ge^2Q^2}{A_{N-1}(N-1)(N-2)r^{2(N-2)}},\\
A^0&=\frac{e^2Q}{(N-2)r^{N-2}}\quad\Rightarrow\quad
F^{0r}=\frac{e^2Q}{r^{N-1}}
\end{split}
\end{equation} Here $M,Q,e,G$ are as before and $A_N$ is the area of a unit $S^N$.

The BH (outer) horizon lies at
\begin{equation}\label{garpl}
r_+^{N-2}=\frac{8\pi GM}{(N-1)A_{N-1}}+\sqrt{\bigl( \frac{8\pi GM}{(N-1)A_{N-1}}\bigr)^2-\frac{8\pi Ge^2Q^2}{A_{N-1}(N-1)(N-2)}}.
\end{equation}
Neglecting order 1 coefficients in the subsequent analysis, we write
\begin{equation}\label{tg}
G\sim m_p^{1-N}.
\end{equation}

Since the extremality bound is of the same form in arbitrary spacetime dimensions, the conditions for the discharge of an extremal black hole in $3+1$ dimensions presented in Eq.~\ref{potp}\ suggests that Eq.~\ref{bound}\ generalizes to higher dimensional spacetimes if instead of the gauge coupling $e$ we use the dimensionless gauge coupling
\begin{equation}\label{te}
\tilde{e^2}\equiv e^2 m_p^{N-3}.
\end{equation}
In the following sections, we will show that this is indeed the
conclusion reached by considering semi-classical discharge
processes\footnote{In principal, it could have been the case that in
$N+1$ spacetime dimensions \ref{bound}\ becomes $\tilde{e}>({m\over
m_p})^{f(N)}$ where $\tilde{e}$ is the {\it dimensionless} coupling
and $f(N)$ is some function of the space dimensions that happened to
obey $f(3)=1$.} resulting in the generalized bound, suggested in
\cite{AMNV}\
\begin{equation}\label{gbound}
\boxed{\tilde{e}\geq\frac{m}{m_p}}.
\end{equation}
This discussion is restricted to space-times with $N\geq 3$ because
for $N\leq 2$ there are no BHs in asymptotically flat space.

\subsection{Hot BHs in $N+1$ dimensions}\label{hbhN}

Using the physical criterion of section \ref{hot}\, the BH will start
to discharge thermally in an appreciable manner when the electric potential
at the horizon is of order the rest mass of the lightest charged particle,
$A^0_{|_{Horizon}}\geq m$. We then obtain the condition that the horizon
radius {\it at this time} (denoted by a tilde) is bounded by
\begin{equation}
\tilde{r}_+^{N-2}\leq \frac{e^2Q}{(N-2)m}.
\end{equation}
Demanding the BHs mass {\it at that time} be large enough to allow for a complete discharge $M_{discharge}\geq
Q\cdot m_{light}$ we get
\begin{equation}
Qm\leq M_{discharge}\leq \frac{e^2Q}{Gm}\quad\Rightarrow\quad \boxed{m\leq \tilde{e}m_p}
\end{equation} where we have used the relations Eq.~\ref{te}\ and Eq.~\ref{tg} and the fact that $r_+^{N-2}$ is of order $GM$ as seen from Eq.~\ref{garpl}. This is the bound \ref{gbound}.

\subsection{Cold BHs in $N+1$ dimensions}

Here, generalizing the argument of section \ref{cold}, the BH will start to discharge appreciably when
$\frac{1}{e^2}\vec{E}^2\cdot\lambda_c^N\geq m$ yielding the
inequality
\begin{equation}\label{line1}
\frac{e^2Q^2}{r_+^{2(N-1)}}\cdot \frac{1}{m^N}\geq m\quad\Rightarrow\quad (r_+m_p)^{N-1}\leq \tilde{e}Q({m_p\over m})^{\frac{N+1}{2}}.
\end{equation}
Rewriting \ref{garpl}\ in dimensionless units and ignoring all order $1$ numbers gives
\begin{equation}\label{simrp}
(r_+m_p)^{N-2}=\frac{M}{m_p}+\sqrt{(\frac{M}{m_p})^2-\tilde{e}^2Q^2}.
\end{equation} Therefore \ref{line1}\ reads
\begin{equation}\label{line2}
 \frac{M}{m_p}+\sqrt{(\frac{M}{m_p})^2-\tilde{e}^2Q^2}\leq (\tilde{e}Q)^{\frac{N-2}{N-1}}({m_p\over m})^{\frac{(N+1)(N-2)}{2(N-1)}}.
\end{equation} Rearranging and squaring we get
\begin{equation}\label{line3}
(\tilde{e}Q)^{2\frac{N-2}{N-1}}({m_p\over m})^{\frac{(N+1)(N-2)}{(N-1)}}-2(\tilde{e}Q)^{\frac{N-2}{N-1}}({m_p\over m})^{\frac{(N+1)(N-2)}{2(N-1)}}\cdot\frac{M}{m_p}+ (\tilde{e}Q)^2\geq 0.
\end{equation} Demanding again that $M\geq Qm$, and dividing the inequality by $Q^2$ we get
\begin{equation}\label{line4}
\tilde{e}^{2\frac{N-2}{N-1}}Q^{-\frac{2}{N-1}}({m_p\over m})^{\frac{(N+1)(N-2)}{(N-1)}}-2\tilde{e}^{\frac{N-2}{N-1}}Q^{-\frac{1}{N-1}}({m_p\over m})^{\frac{N(N-3)}{2(N-1)}}+ \tilde{e}^2\geq 0.
\end{equation}
Now note that for $N\geq 3$ the fraction $1/2\leq
\frac{N-2}{N-1}\leq 1$ and therefore the inequality has the same
qualitative features as the condition $f(\tilde e)\geq 0$, where
\begin{equation}
 f(\tilde e)=\tilde e^{3/2}-\tilde e^{3/4}+\tilde e^2.
\end{equation}
$f(\tilde{e})$ is graphed in Figure 1.
\begin{figure}
 \centering
 \includegraphics[width=5cm,height=3cm,bb=91 3 322 146]{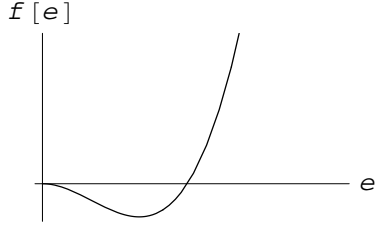}
 \caption{The qualitative behavior of the bound.}
 \label{fig:1}
\end{figure}
Therefore, we can get a bound by finding the non trivial zero
$\tilde e_0$ of $f$, and demanding that $\tilde e\geq \tilde e_0$.
It is useful to look for the strictest bound as a function of $Q$.
We thus choose the equal sign in \ref{line4}, which defines $\tilde
e_0$. We then extremize with respect to Q. Because only two terms
depend on $Q$ we get the following relation for the charge $\hat Q$
that gives the strictest bound
\begin{equation}\label{line5}
(\widehat Q)^{-\frac{1}{N-1}}=\tilde
e_0^{-\frac{N-2}{N-1}}({m_p\over m})^{-\frac{N^2+N-4}{2(N-1)}}.
\end{equation} Plugged back into \ref{line4}\ this gives
\begin{equation}
\boxed{\tilde e_0^2\geq ({m_p\over m})^2},
\end{equation}  which is exactly the expected bound \ref{gbound}.

We conclude that to avoid the problems discussed in section \ref{youone}, in arbitrary spacetime dimensions,
one is universally faced with \ref{gbound}\ as a lower bound on gauge interactions in
the presence of gravity.

\section{String theory consistency checks for $N>3$}\label{highdim}

Having established the bound \ref{gbound}\ in all dimensions $N \geq
3$, by a semiclassical gravitational argument, it is interesting to
perform some checks of it in the context of string theory. We need
to look for situations where a $U(1)$ gauge field and gravity
coexist\footnote{Note that (except for $D9$ branes) the worldvolume
theory of Dp branes in superstring theory is made up only of open
string modes and thus does not contain gravity.}. We will show
agreement with the bound in the following 3 cases
\begin{itemize}
    \item Type-I superstring compactifications.
    \item $D0$ branes in type IIA superstring theory.
    \item $Dp$ branes in type II with a compactified worldvolume.
\end{itemize}

\subsection{Type I String Theory on $\field{R}^{1,9-d}\times \mathcal{M}^{(d)}$}

Consider type-I superstrings compactified down to $10-d$ spacetime
dimensions on a $d$ dimensional compact manifold of volume
$V^{(d)}\sim R^d$. Suppose the compactification breaks $SO(32)$ down
to a $U(1)\times \dots$. For $R > l_s$, the lightest charged
particles under this $U(1)$ subgroup are the off diagonal $SO(32)$
gauge bosons\footnote{In a T-dual picture, those are string states
connecting the separated $D8$ brane to the remaining stack.} whose
mass is set by the KK scale $R$. As long as $R\gg l_s$ the mass of
the lightest charged particle is $m_{light}\sim\frac{1}{R}$. When
$R<l_s$ there are stringy excitations that are charged under this
$U(1)$ and are lighter than the KK modes. Therefore, it is true in
this case that
\begin{equation}\label{mlig}
m_{light}\leq m_s.
\end{equation}
Having said that we can now use the following relations
\begin{equation}
\begin{split}
m_s^4&=g_sm_{(10)}^4\\
m_{(10)}^8R^{d}&=m_{(10-d)}^{8-d}\\
\frac{1}{g_{(10-d)}^2}&=\frac{V^{(d)}}{g_{(10)}}\\
[g_{(10-d)}]_m&=\frac{d-6}{2}\quad\Rightarrow\quad \tilde{g}_{(10-d)}=g_{(10-d)}\cdot m_{(10-d)}^{\frac{6-d}{2}}\\
g_{(10)}&\sim \sqrt{g_s}m_s^{-3}\\
\end{split}
\end{equation} where we denote by $m_{(d)}, g_{(d)}$ the Planck mass and the YM coupling in $d$ spacetime dimensions.
It is straightforward to conclude from these equations that
\begin{equation}\label{der}
g_{(10-d)}=g_{(10)}R^{-\frac{d}{2}}=\frac{\sqrt{g_s}}{m_s^3}\cdot\frac{m_{(10)}^4}{m_{(10-d)}^{\frac{8-d}{2}}}=\frac{m_s}{\sqrt{g_s}}\frac{1}{m_{(10-d)}^{\frac{6-d}{2}}}\cdot\frac{1}{m_{(10-d)}}
\end{equation}
Thus,
\begin{equation}
\tilde{g}_{(10-d)}\cdot m_{(10-d)}=\frac{m_s}{\sqrt{g_s}}\geq m_s>m_{light}
\end{equation} confirming the bound \ref{gbound}. These inequalities hold whenever $g_s< 1$.
For $g_s>1$ one would need to go to a dual weakly coupled picture.
Note that when $R \rightarrow 0$ we have a dual picture of
separated D branes.  Thus, at weak coupling \ref{gbound}\ holds in
compactifications of type-I superstrings in arbitrary spacetime
dimensions, as expected from the general analysis in section
\ref{gengen}.

\subsection{$D0$ branes in flat $\field{R}^{1,9}$}

$D0$ branes are the lightest charged objects under the RR 1-form
potential in type IIA superstring theory. In this case
\begin{equation}
\begin{split}
m_s^4&=g_sm_{(10)}^4\\
\frac{1}{g^2}&=\frac{1}{l_s^6}\\
[g]_m&=-3\quad\Rightarrow\quad \tilde{g}=(\frac{m_{(10)}}{m_s})^3\\
\end{split}
\end{equation}
and it is trivial to check that indeed \ref{gbound}\ is obeyed
\begin{equation}\label{do}
m_{D0}=\frac{m_s}{g_s}=m_s\cdot(\frac{m_{(10)}}{m_s})^4=(\frac{m_{(10)}}{m_s})^3\cdot
m_{(10)}=\tilde{g}\cdot m_{(10)}.
\end{equation} In fact, this must have been true because the $D0$ branes are ${1\over
2}$BPS. This case gives a check of \ref{gbound}\ in flat
$\field{R}^{1,9}$

\subsection{Branes and higher dimensional p-form potentials in flat space.}

It is tempting to generalize \ref{gbound}\ also to the case where
the $U(1)$ symmetry is mediated by a p-form gauge field with $p>1$
so that the lightest charged objects are not ``particles'' (1
dimensional world volume) but ``branes''. One needs to figure out
how exactly to generalize \ref{gbound}\ to this case because
branes have an infinite mass (but a finite tension). In fact, for
branes the semi-classical decay arguments we used in this paper are void because, having infinite mass,
{\it they will not be emitted by BHs at all.} One way to get a
meaningful set up is to compactify the $p$ spacelike directions
along the brane's worldvolume so that it becomes an effective
particle in the $8-p$ dimensional transverse space. This particle
is the lightest state charged under the RR gauge field.

Let us then compactify the spacelike directions of the brane on a
compact manifold of volume $V^{(p)}\sim R^p$. In this case
\begin{equation}\label{comdp}
\begin{split}
\frac{1}{g^2}&=\frac{1}{l_s^{6-p}R^p}\\
[g]_m&=\frac{p}{2}-3\quad\Rightarrow\quad \tilde{g}=g\cdot m_{(10-p)}^{3-\frac{p}{2}}\\
m_{(10-p)}^{8-p}&=\frac{R^p}{g_s^2l_s^8}\\
m_{Dp}&=\frac{R^p}{g_sl_s^{p+1}}\\
\end{split}
\end{equation}

Therefore we write
\begin{equation}\label{derdpdp}
\tilde{g}\cdot m_{(10-p)}=g\cdot
m_{(10-p)}^{4-\frac{p}{2}}=l_s^{3-\frac{p}{2}}R^{p\over
2}\cdot\frac{R^{p\over 2}}{g_sl_s^4}=m_{Dp}
\end{equation} where again because of the BPS property we get an
equality.

\section{Other types of symmetries}\label{nags}

\subsection{Non abelian gauge symmetry}

At first sight, the extension of the AMNV bound to the gauge
coupling of Type I string theory in ten dimensions seems to fail.
If we ignore the gauge supermultiplet, the lightest state charged
under the gauge group $SO(32)$ is a perturbative string state. A
naive application of the AMNV bound might lead us to expect that
\begin{equation}\label{typei}
m_s\leq \tilde g_{YM} m_p.
\end{equation}
The relation between the 10 dimensional string and Planck scales is
given by
\begin{equation}\label{msmp}
m_s^4=g_sm_{(10)}^4
\end{equation} where $g_s\sim e^{\phi}$ is the closed string coupling\footnote{A handle costs $g_s^2$.}.
The 10 dimensional YM coupling is irrelevant and is given by
\begin{equation}
g_{YM}^2\sim g_sm_s^{-6}\quad\Rightarrow\quad \tilde
g_{YM}^2=\frac{1}{\sqrt{g_s}}.
\end{equation} Plugging this back into \ref{msmp}\ gives
\begin{equation}
m_s=\frac{m_{(10)}}{\tilde g_{YM}}
\end{equation} which is in some sense the ``opposite'' of \ref{gbound}.

In fact, there is no good reason to ignore the gauge multiplet.
The essence
of non-abelian gauge theory is that gauge bosons are charged. In
spacetime dimensions higher than four ($N>3$ in the notations of
this paper) the bound \ref{gbound}\ is satisfied trivially for any
non abelian gauge group\footnote{Recall that one cannot Higgs the
gauge group in ten dimensional Type I.}. In those dimensions the
gauge coupling is irrelevant, and therefore the IR gauge theory is
free. Thus, the massless gluons are the lightest charged particles.
In the language of section 1, $m=0$, and so any gauge coupling is
allowed by the bound.   Of course, in {\it e.g.} ten dimensional
Type I string theory, there {\it is} a relation between the gauge
and gravitational couplings, related to the low cutoff scale $m_S$.
However, it cannot be phrased in terms of the mass of the lightest
charged particle in low energy effective field theory, nor derived
from the low energy arguments we have presented here.

In four spacetime dimensions many {\it asymptotically free gauge
theories}  confine and there are no global charges. The nearest we
can come to an analog of the AMNV bound is the statement
\begin{equation}
\Lambda_{QCD}\leq m_p ,
\end{equation} which we think would be accepted by any effective
field theorist.    Since there is no surprise here, the non-abelian
analog of the AMNV bound again appears to be of little utility.

Similar remarks are valid for the other phases of four dimensional
gauge theory.   In a non-Abelian Coulomb phase, there are not really
any particles, but certainly the mass gap in the charged sectors
vanishes. In the Higgs phase there are really no charged particles
either. All particles are created from the vacuum by gauge
invariant operators. However, if one looks at the broken non-abelian
symmetry which acts on gauge invariant states, the gauge bosons
themselves have masses
$$m_W \sim e \cdot v \leq e \cdot m_P ,$$ so an analog of the AMNV bound is the
statement that the Higgs VEV is less than the Planck scale. Again, short of the fact that the mass of the W-boson is much lighter than the Planck mass at weak coupling, the effective field theorist encounters no surprises.

\subsection{Discrete symmetries.}

The problems with entropy considerations, which form the physical
basis for the analysis done in this note, arise because the $U(1)$
charge can assume arbitrarily large values. For a general symmetry
group, particle states sit in irreducible representations of the
symmetry group. {\it Discrete finite groups} have only finitely many
irreps and the sum of the squares of their dimensions add up to the number of elements in the group. 
Thus, one can potentially run into trouble with discrete symmetry groups if
they have infinite order, or if in some sequence of models their order can be
increased without bound.

String theory is full of infinite discrete groups: the duality
groups of super Poincare invariant compactifications.   However,
these groups are effectively spontaneously broken.   A generic
transformation changes the value of the moduli and {\it does not act on
the particle states} of a given scattering matrix.   The subgroup that does act on particle states is, in all known examples,
finite.  There are simple mathematical explanations of this fact
for all known duality groups, but our considerations suggest a
general physical reason.
Particles of finite mass $m$ can not sit in irreps of arbitrarily large size without violating the covariant entropy bound.

\section{Conclusions}\label{conc}
We have presented a semi-classical derivation of the AMNV bound on
the couplings of $U(1)$ gauge fields in arbitrary spacetime
dimensions that allow for BH solutions with flat asymptotics. Our
derivation is based on the the requirement that charged black hole
evaporation not lead to contradictions with entropy bounds. We also
investigated an analogous bound for non-Abelian gauge theories,
without finding situations that would be shocking to an effective
field theorist, and made some remarks about the case of discrete
symmetries. Finally, we performed some simple checks of the bounds
for systems involving D-branes in various space-time dimensions.

\subsection*{Acknowledgments}

It is a pleasure to thank Anthony Aguirre, Michael Dine and Howard Haber
for useful discussions. We also thank Nima Arkani-Hamed and Cumrun Vafa for comments on the manuscript. 
This research is supported by DOE grant DE-FG03-92ER40689.

\end{document}